\documentclass[twoside]{article}
\usepackage{fleqn,espcrc2}

\usepackage{epsfig}
\usepackage[figuresright]{rotating}

\newcommand{\AmS}{{\protect\the\textfont2
  A\kern-.1667em\lower.5ex\hbox{M}\kern-.125emS}}

\title{Exclusive channels in $\gamma\gamma$ reactions:
light at the end of the tunnel~?}

\author{M.R. Pennington\address{Centre for Particle Theory, University of Durham, Durham DH1 3LE, U.K.}%
        \thanks{Travel supported by EEC-TMR Network EuroDA$\Phi$NE, Contract No. CT98-0169.}}
       
\begin{document}

\begin{abstract}
 The physics that can be learnt by studying exclusive channels in two photon interactions is recalled. This serves as an introduction to the Exclusive Reaction session of Photon'99. 
\vspace{1pc}
\end{abstract}

\maketitle
\section{WHY EXCLUSIVE CHANNELS}
  Our main source of information about two photon interactions is  the classic Weizs\"acker-Williams process $e^+e^-\to e^+e^- X$, where for exclusive channels  $X$ is
typically two to six mesons or a baryon-antibaryon final state. The bulk of the data
is obtained without tagging. Then, thanks to the behaviour of the photon propagator, the photons are almost on-shell. By factoring off the known $e^+e^-\gamma$ vertices, we learn about real $\gamma\gamma$ reactions --- see for example [1]. With tagging one can, of course, also investigate important $\gamma^*\gamma$ collisions. I will not review these, but rather refer to the later talk of Bernard Pire~[2]. We can also study the time reversed process, as in E835 at Fermilab, by looking at $p{\overline p}\to\gamma\gamma$~[3].

\begin{figure}
\begin{center}
~\epsfig{file=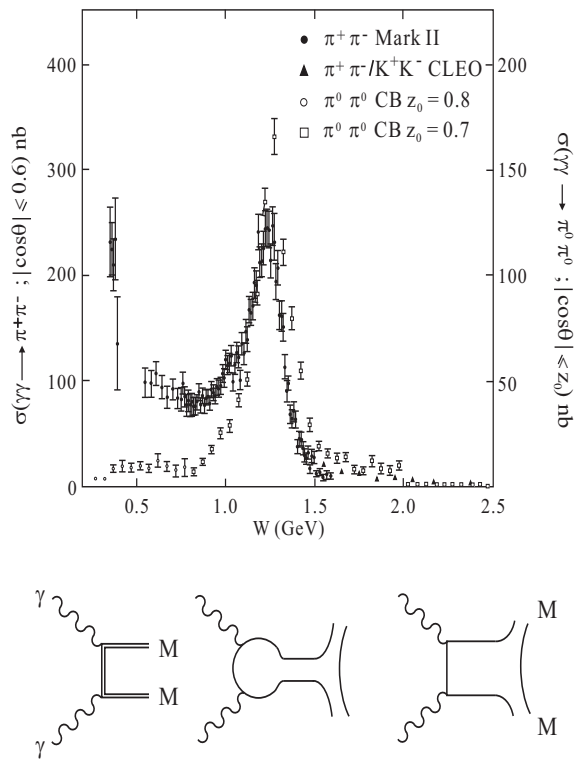,height=10.cm,width=7.5cm}
\vspace{-6mm}
\caption{
{Integrated cross-section for $\gamma\gamma\to\pi\pi$ as a function of c.m. energy $W$ from Mark II~[4], Crystal Ball (CB)~[5] and CLEO~[6].
Below are graphs describing the dominant dynamics in each kinematic region,
as discussed in the text. }} 
\end{center}
\vspace{-8mm}
\end{figure}
  
Now consider the $\gamma\gamma\to MM$ cross-section as a function of the $\gamma\gamma$ c.m. energy $W$. A typical example is shown in Fig.~1. There one sees the cross-section rise from threshold, then have structure and subsequently decline. This cross-section naturally divides into three kinematic regions which correspond to three different dynamical regimes.
In each case the photon couples to the electric charge of a point-like object, but what it sees as point-like  changes with energy. At low energy, close to threshold, the photon has long wavelength and sees the whole of the final state hadron and couples to its electric charge. This teaches us about
{\it hadron dynamics}. As the energy increases and the wavelength of the photon shortens, it sees the charged components of the hadron, the constituent quarks. Coupling to them, it causes them to resonate, so we can learn about {\it resonance dynamics}. Lastly, as the energy rises still further, the photon sees charged point-like objects inside the constituent quarks, the current quarks, and we can learn about {\it quark dynamics}. The extent  of the three kinematic regions depends on the final state.
For pions, the three regions are well-separated, for kaons the near threshold region is foreshortened and the resonance regime more structured. In all three cases, the dynamics is governed by QCD. Region 3 is the perturbative--non-perturbative interface, while the lower two depend wholly on the strong physics aspects of QCD. For charmonium and charmed particle production, regions 2 and 3 merge with each other, both being amenable to perturbative treatment.

  Let us discuss region 3, the higher energy regime first, as that is the easier to explain. Above a few GeV, when two photons collide in the centre-of-mass frame, they deposit all of their energy in a region the size of a fraction of a fermi. This creates a $q{\overline q}$ pair moving back-to-back. These radiate soft gluons, until they reach separations of the order of a fermi, when the
quarks and gluons fragment. This produces two back-to-back jets of hadrons that dominate the inclusive high energy cross-section. To produce an exclusive final state like two mesons, the initial $q{\overline q}$ pair must radiate at least one hard gluon, which in turn creates another $q{\overline q}$ pair, moving back-to-back parallel to the initial quarks, so that these can get together to form a meson.  It was Brodsky and Lepage~[7], who long ago, recognised that such processes can be divided into a short-distance part governing the emission of the hard gluon and a long distance component determined by the wavefunction of the hadron final state. Thus they predicted that the differential cross-section for
$\gamma\gamma\to\pi\pi$ should be given by
\begin{equation}
\frac{d \sigma}{d \cos \theta}\ (\gamma\gamma\to\pi^+\pi^-) \simeq \frac{8\pi\alpha^2}{W^2}\;\frac{F_{\pi}^2 (W^2)}{\left(1-\cos^2 \theta\right)^2}\; ,
\end{equation}
where the wavefunction of the pion is related to the pion's electromagnetic formfactor $F_{\pi}(W^2)$. This prediction (Fig.~2) is in reasonable agreement with data from Mark II~[1,8], TPC/Two-Gamma~[1,9], CLEO~[6], et al. To form baryons in the final state, a further hard gluon has to be radiated, to produce yet another back-to-back $q$ and ${\overline q}$. It was Farrar, Maina and Neri~[11], who first extended this discussion to $p{\overline p}$ production, by building in the proton's three quark
wavefunction of Chernyak and Zhitnitsky~[12]. As seen in Fig.~2, this makes a prediction  well below the data. Though baryons are built of three valence quarks, they spend part of their time in a quark-diquark configuration. Perhaps baryons
are produced by hard gluons  creating just one diquark-antidiquark,
instead of 
two $q{\overline q}$ pairs. This gives the predictions~[13] shown in Fig.~2, for both the energy and scattering angle dependence, in far better agreement with data.  Unfortunately, this detailed modelling is at variance with more recent CLEO data on $\Lambda {\overline \Lambda}$ production~[14].

\begin{figure}
\begin{center}
~\epsfig{file=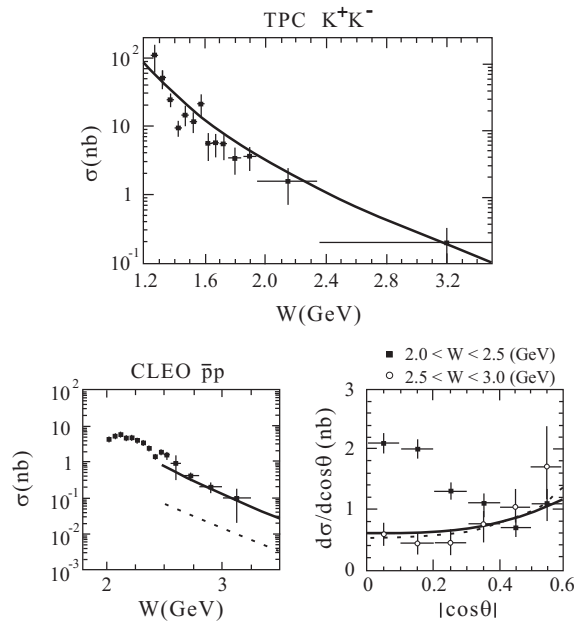,height=8.2cm,width=7.5cm}
\vspace{-13mm}
\caption{
{Integrated cross-section for $\gamma\gamma\to K^+K^-$ as a function of c.m. energy $W$ from TPC/Two-Gamma~[9] compared to the predictions of Brodsky and Lepage~[7]. Below is the cross-section for
$\gamma\gamma\to p{\overline p}$ from CLEO~[10] as a function of $W$ on the left and as a function of the c.m. scattering angle $\theta$ on the right, in two energy bins compared with the prediction of Farrar {\it et al.}~[11] (dashed line) and of the diquark model of Anselmino {\it et al.} and Kroll {\it et al.}~[13] (solid line).} } \end{center}
\vspace{-13mm}
\end{figure}

\section{RESONANCE DYNAMICS}

  If we now look at the  energy region of $W < 2$ GeV, we see  distinct resonance structures, for instance in $\gamma\gamma\to K^+K^-$~[15] in Fig.~3.
There three well-known spin two resonances, the $f_2$, $a_2$ and $f_2'$ appear.
Tensor mesons always appear strongly in vector-vector interactions. 
In the $K{\overline K}$ channel, the $f_2$ and $a_2$ overlap, so the data display an interference effect. We know these mesons belong to an almost ideally mixed quark model multiplet, but nevertheless we can use the two photon data to test this. The photons couple to the electric charges of the quarks. Thus the two photon amplitude, Fig.~4, measures their mean charge squared and the cross-section (and radiative width) gives the square of this.
\begin{figure}
\begin{center}
~\epsfig{file=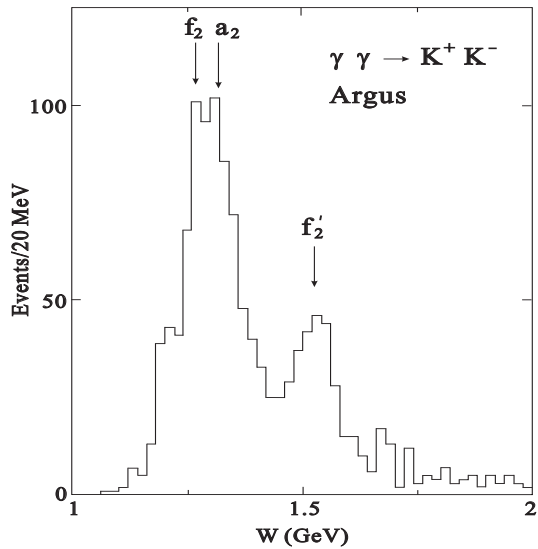,height=7.2cm,width=7cm}
\vspace{-8mm}
\caption{
{$\gamma\gamma\to K^+K^-$ events as a function of c.m. energy $W$ from Argus~[15] in the resonance region, i.e. $W < 2$ GeV.} }
\vspace{-8mm} 
\end{center}
\end{figure}
Thus 
\begin{equation}
\Gamma(R\to\gamma\gamma)\,=\,\alpha^2 \, \langle e_q^2 \rangle^2\;\Pi_R\; ,
\end{equation}
where $\Pi_R$ is the probability that the quarks annihilate. It is natural to assume for such a multiplet that the $\Pi_R$ are the same for each member. Then the quark model simply predicts
\begin{eqnarray}
\Gamma(f_2\to\gamma\gamma)\, :\, \Gamma(a_2\to\gamma\gamma)&:&\Gamma(f_2'\to\gamma\gamma)\,\cr
&&\cr
&=& 25\, :\, 9\, :\, 2\; ,
\end{eqnarray} 
which is sensitive to small deviations from ideal mixing.
Experiment confirms the closeness to this  pattern with $25\,:\, (10 \pm 2)\,:\,(1 \pm 0.2)$.

  So much for relative rates, but what about absolute predictions for
the radiative widths. For these one needs a modelling of the dynamics.
The simplest to imagine is by analogy with positronium decay. We begin
with a fermion-antifermion bound state with definite quantum numbers
$^{2S+1} L_J$. This decays by one fermion exchange, as shown in Fig.~4a.
This is readily calculated as shown by Godfrey and Isgur~[16],
Ackleh and Barnes~[17], Klempt et al.~[18], Resag and M\"unz~[19], Schuler et al.~[20],  and many others.
The  result simplifies for heavy fermion systems, where one can safely take the non-relativistic limit. Then the bound states formed by the interquark
potential have radial wavefunctions $\psi(r)$ that can be straightforwardly calculated using the 
Schr\"odinger equation. This gives, for instance, for  spin singlet states
\begin{figure}[t]
\begin{center}
~\epsfig{file=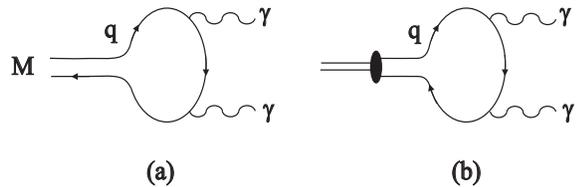,height=2.4cm,width=7.5cm}
\vspace{-10mm}
\caption{
{Meson decay to $\gamma\gamma$ by the exchange of (a) a free fermion, (b) a confined fermion. In both cases, graphs with the two photons interchanged are required by gauge invariance. }} 
\end{center}
\vspace{-10mm}
\end{figure}
\begin{equation}
\Gamma\left( ^1 J_J\to\gamma\gamma\right)\,=\, \frac{3\ \langle e_q^2\ \rangle^2\
\alpha^2}{m_q^{\;2J+2}}\, \mid \psi ^{(J)} (0)\mid^2\; .
\end{equation}
This involves the $J$th derivative of the wavefunction at the origin, which is computable for charmonium, assuming the form of the interquark potential, Coulombic plus linear confinement, for example.
Note that the mass of the fermion enters, but not the mass of the bound state. While for heavy flavour systems it may seem eminently sensible to assume the {\it effective} mass of the quark
is half that of the bound state, for light quark mesons, this is not so obvious.  The relativistic analogue of Eq.~(4) is given in [17] for instance. From this one would expect the radiative width of the
$\pi^0$, $\eta$ and $\eta'$ to be roughly the same, taking into account merely differences in their $\langle e_q^2 \rangle^2$, since they are made of quarks with constituent masses of 330--450 MeV (depending on their strange quark
component). In contrast, experiment gives
\begin{eqnarray}
\nonumber
\Gamma(\pi^0\to\gamma\gamma)\; :\; \Gamma(\eta\to\gamma\gamma)&:& \Gamma(\eta'\to\gamma\gamma)\cr
&&\cr
 &\sim & \; 1\,:\,60\,:\,500.
\end{eqnarray}
The recipe for rescuing this potential disaster was suggested long ago by 
Hayne and Isgur~[21]. They proposed multiplying by a factor of the mass of the meson cubed. This has been justified by Ackleh and Barnes~[17] by noting that
if one considers the point interaction of a pseudoscalar with two photons,
then the gauge invariant form of the interaction Lagrangian is
\begin{equation}
{\cal L}_I\;=\;\frac{1}{2}\, g\,\phi\, {\cal F}^{\mu\nu}\,{\widetilde{\cal F}_{\mu\nu}}\; .
\end{equation}
This gives $\Gamma(R\to\gamma\gamma) \sim  g^2 M_R^3$, where $R$ is a pseudoscalar meson and $M_R$ its mass. If we take it that what we have calculated by considering the fermion exchange graphs of Fig.~4a is some effective coupling $g$, then this has to be
corrected by 
 $(M_R/M_{ref})^3$, where $M_{ref}$ is typically 1 GeV for light hadrons~[21,16,17]. Whilst such factors bring agreement with experiment for the $\pi^0$ and $\eta$, they also affect predictions for the $\eta_c$, $\eta'_c$ too,
because these mass terms 
appear cubed. In Tables 1,2  we show the predictions for radiative widths for
different modellings for a handful of charmonium and light quark states. We will hear at this meeting, new results to compare with these. The absolute rates for the $f_2$ and $a_2$ look good. Of course, the quark model not only predicts ground states, but radial excitations too.
Whilst some, like the $\pi(1300)$, have been seen in hadronic reactions for almost twenty years,
it is only very recently that a candidate, the $a_2(1750)$, has been observed in two photon channels by L3~[22] at LEP. For light quark systems, in particular, what is needed are  calculations beyond the free quark exchange model.
Quarks are confined. There are no poles in their propagators and the bound state dynamics more complicated (Fig.~4b). Meaningful
 non-perturbative computations should be possible soon.
\begin{table}[htb]
\caption{Radiative widths of some charmonium states predicted by variations on the modelling~[16,17,19] described in the text compared with the PDG'98 average~[23] ---
in the case of the $\chi_{c0}$ this is just one experiment.}
\label{table:1}
\renewcommand{\arraystretch}{1.2} 
\vspace{3mm}
\begin{tabular}{@{}lllll}
\hline
$\Gamma(\gamma\gamma)$ & Godfr. & Ackleh- & Resag & PDG
 \\[-2pt]
keV & -Isgur & Barnes & -M\"unz & '98\\ 
\hline
$\eta_c$   & $\;3.69$          & $\;4.80$ & $\;3.82$ & $\;7.5\,  ^{+1.6}_{-1.4}$ \\
$\chi_{c0}$                & $\;1.29$ & $\;1.56$ & $\;1.62$ & $4.0 \pm 2.8$ \\
$\chi_{c2}$  & $\;0.46$  & $\;0.56$ & $\;0.60$ & $0.36 \pm 0.17$  \\
\hline
\end{tabular}\\
\vspace{-5mm}
\end{table}

\begin{table}[htb]
\caption{Radiative widths of some typical light meson states as predicted by two models~[18,19], compared with the PDG'98 averages~[23].}
\label{table:2}
\newcommand{\m}{\hphantom{$-$}}
\newcommand{\cc}[1]{\multicolumn{1}{c}{#1}}
\renewcommand{\arraystretch}{1.2} 
\vspace{2mm}
\begin{tabular}{@{}lllll}
\hline
$\Gamma(\gamma\gamma)$ & Klempt & M\"unz & 
PDG \\[-2pt]
keV & ~et al. & & '98 \\
\hline
$\pi(138)*10^3$    & $\;4.23$ & $\;5.07$ & 
$7.74 \pm 0.55$  \\
$\eta(547)$    & $\;0.208$ & $\;0.014$ & $0.46 \pm 0.04$ \\
$\eta'(958)$  & $\;2.33$ & $\;0.046$  & $4.27 \pm 0.19$ \\
$f_2(1270)$   & $\;2.04$ & $\;2.5$ & $2.80 \pm 0.40$ \\
$a_2(1320)$   & $\;0.73$ & $\;0.9$ & $1.00 \pm 0.06$ \\
$a_2'(1750)$  & $\;0.37$ & $\;0.38$ & $(0.29 \pm 0.04)/$ \\
 & & & BR($a_2'\to 3\pi$) \\
\hline
\end{tabular}\\
\vspace{-6mm}
\end{table}
\section{GLUEBALL SEARCHES}
  A much advertised role of two photon processes is the light (or rather lack of it) they shine on glueballs. Glueballs are bound states
of the pure glue sector of QCD.  While experiment for the most part studies processes in which quarks are intrinsic, lattice calculations readily investigate the pure glue world.
These find the scalar the lightest, and the tensor the next heaviest, bound states. The last few years have seen a dramatic reduction in the uncertainties on  lattice masses.
The scalar mass is
${\cal M}_0(gg)\;=\; (1730\,\pm\,94) \;{\rm MeV}$,
averaging UKQCD~[24], GF11~[25] and the improved action~[26] results. The tensor is 1.5 times heavier around 2300 MeV. Let me here concentrate on the scalar candidates, and leave the tensor to Hans Paar~[27].

The known lightest scalars are the $f_0(400-1200)$, $f_0(980)$, $a_0(980)$,
$K_0^*(1430)$, $f_0(1500)$ and $f_0(1710)$, to which may be added possible $f_0(1370)$, $a_0(1430)$ discounting a $\kappa(800)$~[23]. 
As has been long known~[28], the nine lightest of these do not fit into an ideally mixed $q{\overline q}$ multiplet.  The $f_0(980)$ is known to couple strongly to $K{\overline K}$. If it were like its vector analogue, the $\phi$, it might be thought to be 
largely $s{\overline s}$ in composition. But then it would be difficult to understand
why, with two strange quarks, it should be lighter than the $K_0^*(1430)$ with one strange quark and yet be degenerate in mass with the isotriplet $a_0(980)$ built only of $u$ and $d$ quarks. This points to the fact that these nine lightest scalars are not simply connected to an underlying $q{\overline q}$ multiplet, in the way that vectors and tensors are.
The $\rho$ and $\phi$, for instance, are predominantly $q{\overline q}$ states.
However, they do contain within their Fock space two meson components,
mainly $\pi\pi$ and $K{\overline K}$, respectively. It is through these meson modes that they decay. For vectors these components are small and do not distort the direct connection with the underlying $q{\overline q}$ states. In contrast, for scalars, these hadron dressings are known to be large. Though the details of their calculation may be model-dependent~[29], it is inevitable that the
$f_0(980)$ and $a_0(980)$ contain large $K{\overline K}$ components that bring them close to $K{\overline K}$ threshold. In lattice-speak, {\it unquenching} matters~[30]. This notwithstanding,
the $f_0(1500)$ and  $f_J(1710)$ have been proposed as glueball candidates,
in keeping with the lattice expectations we have just discussed, for their {\it undressed} masses. Very recently, Minkowski and Ochs~[31] have proposed that it is the broad $f_0(400-1200)$ that is largely gluish. Since any of these states sits amongst the $q{\overline q}$ scalars they inevitably mix with them, whether at the bare or {\it dressed} levels. This is in contrast to the tensor glue candidate, the $\xi(2230)$, that Hans Paar will discuss~[27], which lies 700 MeV above the ground state $q{\overline q}$ multiplet.
  
\begin{figure}[t]
\begin{center}
~\epsfig{file=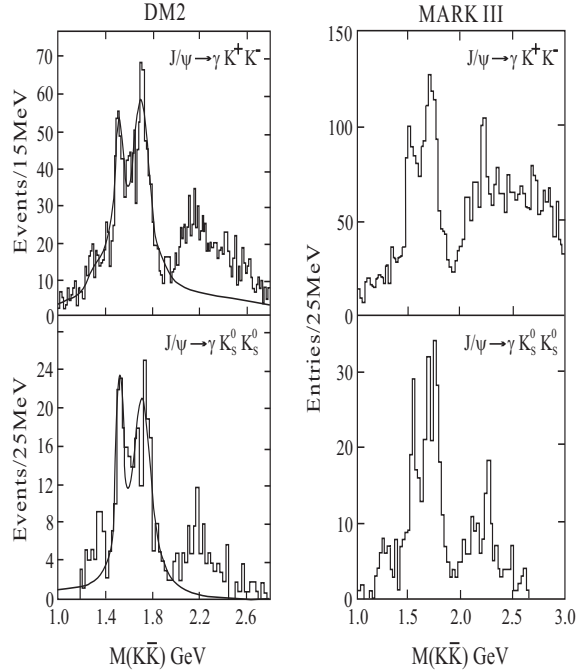,height=8.9cm,width=7.5cm}
\vspace{-11mm}
\caption{
{$K{\overline K}$ mass spectra in $J/\psi$ radiative decay from DM2~[34] and MarkIII~[33] in both the charged and neutral kaon modes.}} 
\end{center}
\vspace{-11mm} 
\end{figure}

The folklore is that glueballs, particularly pure ones, should appear strongly in glue-rich processes like $J/\psi$ radiative decay, yet be dark
to photon interactions. Indeed, the relative rate with which a resonance appears in these two processes defines what Chanowitz called {\it stickiness}~[32].
 We will comment on this below. First we recall, Fig.~5, the $K^+K^-$ and $K_sK_s$ mass distributions in $J/\psi\to\gamma K{\overline K}$ from Mark III~[33] and 
DM2~[34]. We see a large signal in the 1700 MeV region, originally called the $\theta$, now the $f_J(1710)$, which is a mixture of spin 0, 2. A smaller signal seen by Mark III, and by BES~[35],
 is the
$\xi(2230)$. These are potentially glueball candidates. But in all these experiments the
well-known tensor meson, the $f_2'(1525)$ is also observed, Fig.~5.
Clearly {\it glue-rich} processes produce conventional $q{\overline q}$ states, as well as putative glueballs, as indicated in Fig.~6. How do we tell the difference?

\begin{figure}[t]
\begin{center}
~\epsfig{file=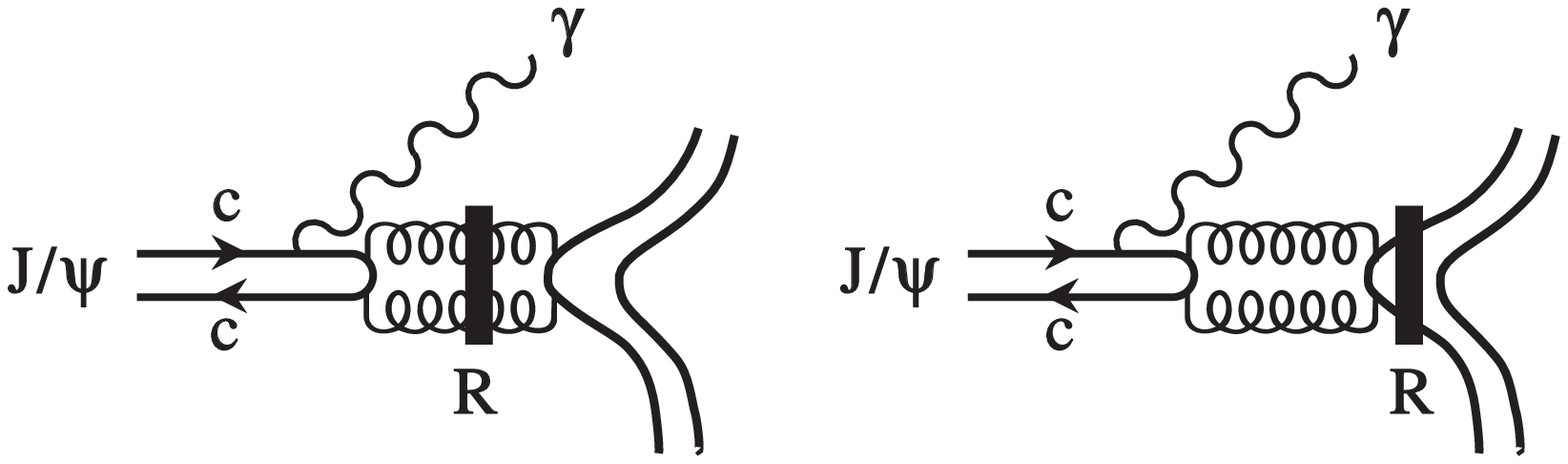,height=3.3cm,width=7.5cm}
\vspace{-12mm}
\caption{
{Resonance formation in $J/\psi$ radiative decay to two mesons. Either the glue can resonate or the $q{\overline q}$ can. The \c Cakir--Farrar method tests the difference~[36,37].}} 
\end{center}
\vspace{-12mm} 
\end{figure}

  \c Cakir and Farrar~[36] thought of a very nice test, based on perturbative ideas of gluon interactions, which Close, Farrar and Li~[37] have applied to $J/\psi$ radiative decays. By comparing resonance production  in exclusive channels to the inclusive rate, one can deduce what might be regarded as the branching ratio of the resonance to two gluons. For a glueball this would be near 100\%, while for a $q{\overline q}$ meson this would involve the gluons creating quarks
and so be suppressed by $\alpha_s^2$, Fig.~6.  Applying this test~[37,38] to the $f_2$ and $f_2'$ gives them a gluonic branching ratio of $\sim 20\%$.
When applied to the $f_J(1710)$ with the PDG averaged result~[23] for $J/\psi$ decay, one finds that if its spin is 2, then the gluonic branching ratio is 15--30\%,
while if it has spin 0 the branching ratio is 50--100\% --- good news for scalar glueball hunters. However, BES~[35] claims to have measured the decay rates for separated $f_2(1696)$ and $f_0(1780)$ components, which implies the latter has a gluonic branching ratio of less than 10\%. This discrepancy needs to be resolved. Unfortunately, similar disparities arise for the $f_0(1500)$ in $J/\psi$ radiative decays, which is only \lq\lq observed'' in its  $4\pi$ decay mode --- see [39].

  To deduce the radiative widths of resonances, and hence have another handle on their composition, requires a clean separation of data into spin
components. This is particularly true for scalars, which lie underneath rather large tensor signals dominating Fig.~1, for example.
If one had data on interactions with polarized photons and complete angular coverage of the hadron final state, this would be relatively straightforward.
However, we have no polarization information and coverage of 80\% of the angular range in $\cos\theta$ at best. It is then only for the $\pi\pi$ final state that one can find sufficient other constraints to make a partial wave separation possible.  $\gamma\gamma\to\pi\pi$ is dominated by the Born term at low energies (Fig.~1), modified by calculable final state interactions.
  In addition unitarity requires that the $\gamma\gamma$ process is related to hadronic reactions.
For instance, the partial wave amplitudes ${\cal F}$ (with definite spin, helicity and isospin) for $\gamma\gamma\to \pi\pi$ are related to 
the corresponding amplitudes ${\cal T}$ for $\pi\pi\to\pi\pi$, $\pi\pi\to K{\overline K}$, etc.,  by~[40]
\begin{eqnarray}
&&\hspace{-9mm}{\cal F}(\gamma\gamma\to \pi\pi)\cr
&=&\rho_1\ {\cal F}^*(\gamma\gamma\to\pi\pi)\
{\cal T}(\pi\pi\to\pi\pi)\cr
&+& \rho_2\ {\cal F}^*(\gamma\gamma\to K{\overline K})\
{\cal T}(\pi\pi\to K{\overline K})
+ \cdots \ ,
\end{eqnarray}

\begin{table}[b]
\vspace{-3mm}
\caption{Radiative widths of states contributing to $\gamma\gamma\to\pi\pi$ below 1.4 GeV for the two solutions ({\it dip} and {\it peak}) found in the Amplitude Analysis of [41]. }
\label{table:3}
\renewcommand{\arraystretch}{1.5} 
\vspace{2mm}
\begin{tabular}{@{}lllll}
\hline
    & &$\Gamma(\gamma\gamma)$ keV& \\
\hline
solution  & $f_2(1270)$           & $\; f_0(980)$ & $f_0(400/1200)$  \\
\hline
dip                & $\quad 2.64$ & $\quad 0.32$ & $\qquad 4.7$ \\
peak    & $\quad 3.04$  & $0.13-0.36$ & $\qquad 3.0$  \\
                
\hline
\end{tabular}\\
\vspace{-3mm}
\end{table}
\noindent where $\rho_i$ are the appropriate phase-space factors.
Below 1.4 GeV or so, when multipion channels start to become important, just the $\pi\pi$ and $K{\overline K}$ intermediate states are all that need be included. By implementing such constraints, one can make up for the inadequacies of the two photon information by incorporating hadronic scattering data into the codes~[40]. Despite this, such an analysis is only possible if data on both $\pi^+\pi^-$ and $\pi^0\pi^0$ final states are included simultaneously. Using the most recent data (including [4,5]), Boglione and I~[41] have completed such an Amplitude Analysis. This reveals two solutions differentiated by whether the $f_0(980)$ appears as a peak or as a dip, Fig.~7. For each of these solutions, one can deduce the radiative widths of the $f_2(1270)$, $f_0(980)$ and $f_0(400-1200)$.
These are listed in Table~3.
\begin{figure}[t]
\begin{center}
~\epsfig{file=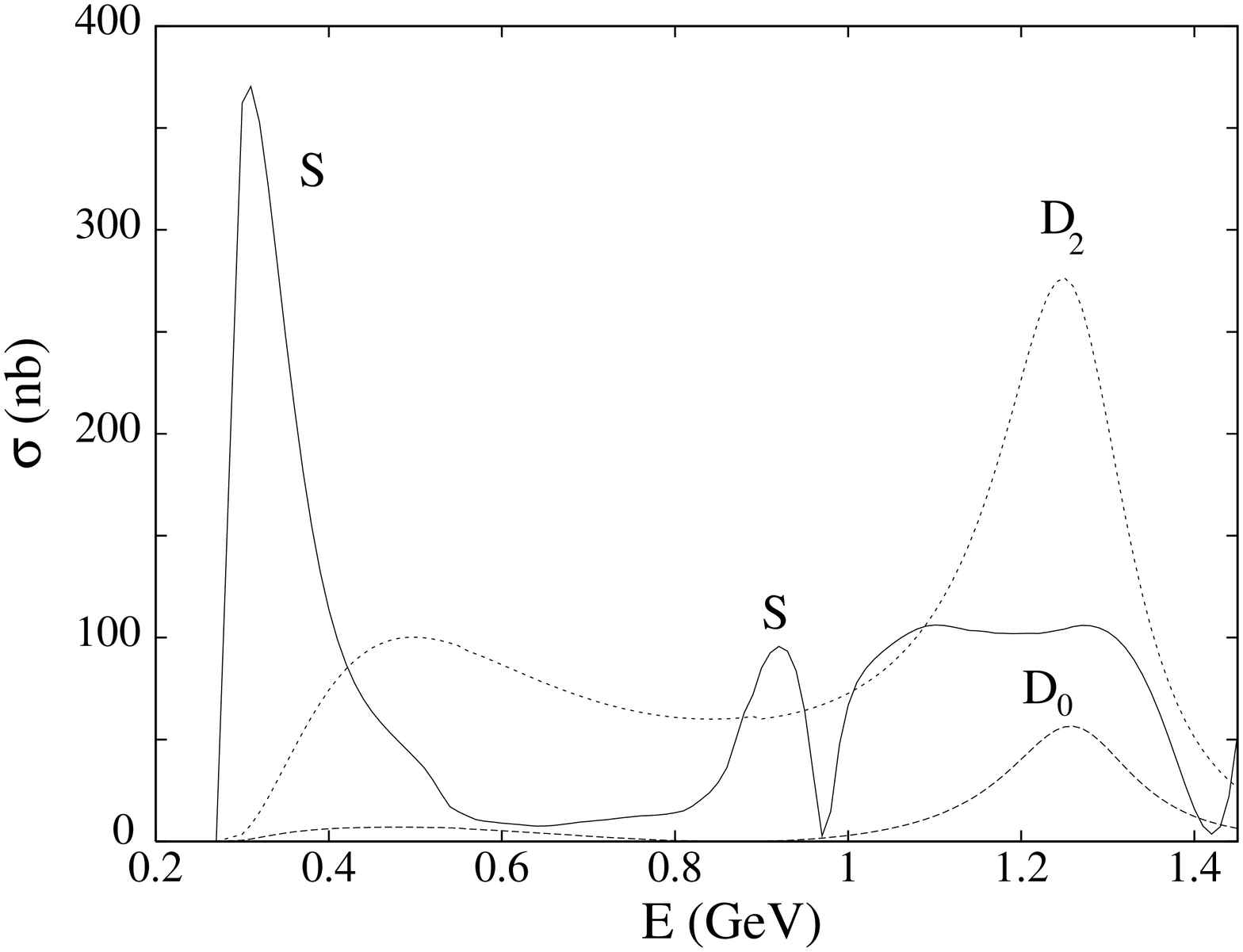,angle=-90,width=7.3cm}
\vspace{-2mm}

~\epsfig{file=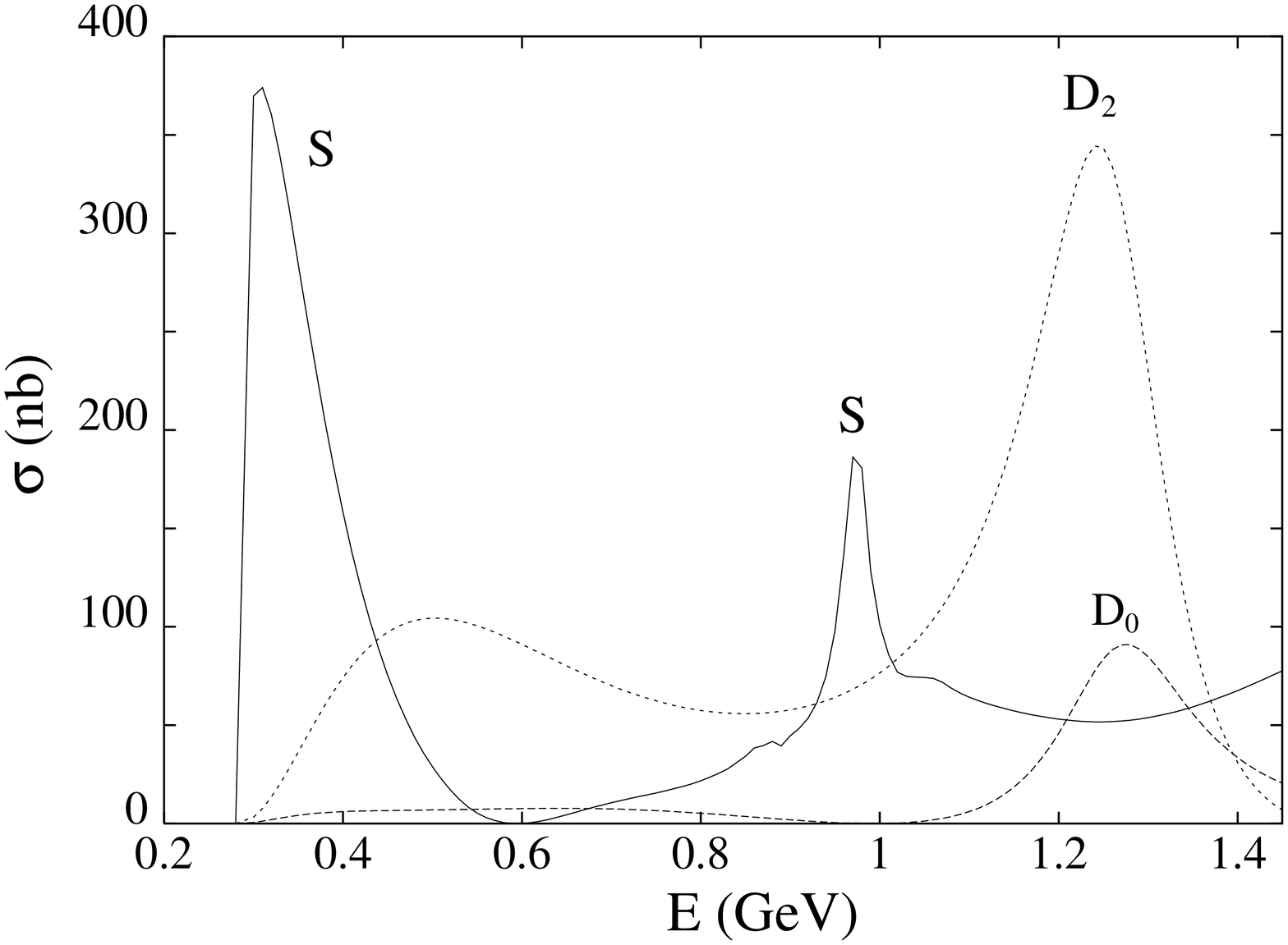,angle=-90,width=7.5cm}
\vspace{-5mm}
\caption{
{Contribution of the individual partial waves, labelled as $J_{\lambda}$, where $J$ is the spin and $\lambda$ the helicity, to integrated $I=0$ $\gamma\gamma\to\pi\pi$ cross-section describing the data of Fig.~1, from the Amplitude Analysis of [41]. The upper figure is for the so called {\it dip} solution, the lower for the {\it peak} solution. The radiative widths of the corresponding resonances are given in Table~3.}} 
\end{center}
\vspace{-10mm}
\end{figure}
 Precision two photon information, differential as well as integrated cross-sections, can distinguish between these solutions.
CLEO took such data some time ago, but unfortunately these have never been 
finalised.  In each solution there is a sizeable $I=0$ $S$--wave signal through the $f_2$ region, required by the difference in shape of the 1270 MeV peak in the $\pi^+\pi^-$ and $\pi^0\pi^0$ modes, seen in Fig.~1.  This $S$--wave signal (Fig.~7) is attributed to the $f_0(400-1200)$ with its radiative width given in Table~3. For the scalars, these widths should be compared with the predictions in Table~4 for different compositions. One sees that for the $f_0(400-1200)$,
the width is consistent with that for a non-strange $q{\overline q}$ state, i.e. $n{\overline n} \equiv (u{\overline u} + d{\overline d})/\sqrt{2}$, while the two photon width of the $f_0(980)$ is compatible with Barnes' calculations~[42] for either an $s{\overline s}$ or $K{\overline K}$--molecular composition.
As alluded to earlier, the $f_0(980)$ is likely to have a more complicated Fock space, in which both $s{\overline s}$ and $K{\overline K}$
components feature strongly~[29,30]. How to calculate the $\gamma\gamma$ coupling of such complexes is as yet unknown~[38].

 \begin{table}[htb]
\vspace{-5mm}
\caption{How the predicted $\gamma\gamma$ width of scalars in the 1--1.3 GeV region depends on their composition~[42,40].}
\label{table:4}
\renewcommand{\arraystretch}{1.5} 
\vspace{2mm}
\begin{tabular}{@{}lllll}
\hline
 keV   &$\qquad n{\overline n}\;$ & $\qquad s{\overline s}\;$ & $\qquad K{\overline K}\;$ \\
\hline
$\Gamma(0^{++}\to\gamma\gamma)$   & $\qquad 4.5$ & $\qquad 0.4$ & $\qquad 0.6$ \\
\hline
\end{tabular}\\
\vspace{-5mm}
\end{table}

What about the higher mass scalars?
In $\pi\pi$ elastic scattering, the $f_0(980)$ and $f_0(1500)$ appear as dips
in the cross-section. It is quite possible that these feed through into the
two photon reaction too (through Eq.~(6)). If this is the case, then the idea of deducing an upper bound on the radiative width of the $f_0(1500)$ by the surplus of events over a smooth background used by ALEPH~[43] to obtain $\Gamma(f_0(1500)\to\gamma\gamma) < 0.17$ keV looks highly questionable.  Only by extending the Amplitude Analysis I have just described to higher $\pi\pi$ masses can one be certain of separating an $S$--wave signal out from under the larger $D$--wave components and so deduce meaningful results for the radiative widths of the
 glueball candidates $f_0(1500)$ and $f_0(1710)$.

Clearly a pure glueball would have a very tiny $\gamma\gamma$ width (coming wholly from higher order graphs in perturbative QCD). Consequently, its {\it stickiness} (defined earlier in this section) would be large. What {\it large} means has always been problematic.
Stickiness~[32] is conventionally normalized so that it is 1 for the $f_2(1270)$, a typical quark model state. However, for its $s{\overline s}$ partner, the $f_2'(1525)$, the stickiness is $\sim 15$. Clearly, the $f_2'$ is not {\it sticky}
in the sense of gluey.
This large value happens because its two photon width is small: a result of it being made of charge 1/3 quarks, while the $f_2$ has both $u$ and $d$ quarks.
The CLEO group~[44] have sought to address this problem.
They define a quantity {\it gluiness}, 
which is roughly equal for the $f_2$ and $f_2'$. However, gluiness~[44] suffers from the drawback that one has to assume in advance what the charged components of the meson are ---
something, of course, we want to use the two photon width to determine.

As already remarked the scalar glueball candidates, primarily the $f_0(1500)$ and $f_0(1710)$,
are unlikely to be pure glue, but inevitably mix with the nearby
$q{\overline q}$ multiplet(s). As soon as this happens, these 
states cease
to have tiny radiative widths, 
but have those typical of $q{\overline q}$ 
scalars, cf.  Table~4.
Stickiness then has much less relevance. The branching ratio to glue of
\c Cakir and Farrar~[36] is far more useful, as is the absolute value of their two photon widths. The predictions for the latter depend very strongly on the assumed mixing scenario as illustrated by comparing the results of
Close, Farrar and Li~[45] with those of Jaminon and van den Bosche~[46]
and with Burakovsky and Page~[47]. These vary by large factors.

The challenge for the near future is for experiment to deliver accurate measurements, from which the two photon widths of the $f_0(1500)$ and $f_0(1710)$ can be deduced, and for theory to make reliable predictions for these same states. Combining this information we can determine the composition of these key hadrons. Only then will we understand the nature of the light scalar mesons, a nature and composition that is intimately tied to the structure of the QCD vacuum --- both $q{\overline q}$ and glue. After decades of promise and years of effort, there may be light at the end of the tunnel.

\vspace{6mm}

\noindent{\bf Acknowledgements}

It is a pleasure to thank Stefan S\"oldner-Rembold and  colleagues for organising a most interesting meeting with such obvious enthusiasm.

\end{document}